\documentclass[aps,pra,twocolumn,showpacs,superscriptaddress]{revtex4-1}
\usepackage{enumerate}
\usepackage{graphicx}
\usepackage{amsmath}
\usepackage{amsfonts}
\usepackage{amssymb}
\usepackage{braket}
\usepackage{gnuplottex}
\usepackage{epsf}
\usepackage{xcolor} 
 
\newcommand{\Tr}{\text{Tr}}

\begin{document}
\today
\title{Proposed neutron interferometry test of Berry's phase for a circulating planar spin}
\author{Erik Sj\"oqvist}
\affiliation{Department of Physics and Astronomy, Uppsala University,
Box 516, Se-751 20 Uppsala, Sweden}
\begin{abstract} 
The energy eigenstates of a spin$-\frac{1}{2}$ particle in a magnetic field confined to 
a plane, define a planar spin. If the particle moves adiabatically around a loop in this 
plane, it picks up a topological Berry phase that can only be an integer multiple of $\pi$. 
We propose a neutron interferometry test of the Berry phase for a circulating planar spin 
induced by a magnetic field caused by a very long current-carrying straight wire 
perpendicular to the plane. This Berry phase causes destructive interference in the 
direction of the incoming beam of thermal neutrons moving through a triple-Laue 
interferometer. 
\end{abstract}
\maketitle
\section{Introduction}
Neutron interferometry has been used to demonstrate experimentally various quantum interference 
effects. These include gravity-induced phase shifts \cite{colella75}, the Aharonov-Casher 
effect \cite{cimmino89}, the scalar Aharonov-Bohm phase shift \cite{allman92}, as well as the 
Levy-Leblond confinement phase \cite{rauch02}. While the neutron spin plays no role or a passive 
role in these experiments, phase effects induced by the evolution of the neutron spin have been 
demonstrated in terms of the geometric phase \cite{wagh97,hasegawa01,klepp08}. 

The Berry phase is the geometric phase arising in cyclic adiabatic evolution of general 
quantum systems \cite{berry84}. In the case of a spin, this phase is proportional to the 
enclosed solid angle $\Omega$ of the slowly rotating magnetic field. Specifically, if the 
spin projection is $m_s$, the Berry phase is $-m_s\Omega$ for a positively oriented path.  
In the case of a neutron, $m_s=\pm \frac{1}{2}$ and the Berry phase can take the values  
$\mp \frac{1}{2} \Omega$. 

If the neutron spin is confined to a plane, forming a planar spin, the enclosed solid angle 
is restricted to an integer multiple of $2\pi$. This implies that the Berry phase picked up 
when the neutron moves around a loop has a topological character as it 
can only be an integer multiple of $\pi$. For an odd multiple, this phase shift results in 
complete destructive interference of certain outcomes. This effect 
has been predicted in scattering of ultraslow neutrons \cite{sjoqvist15}, and in time evolution 
of other two-level quantum systems, such as $E\otimes \epsilon$ Jahn-Teller molecules 
\cite{schon94}, cavity QED systems \cite{larson08}, and degenerate pairs of dark states in 
cold atoms \cite{larson09}. 

The setup proposed in Ref.~\cite{sjoqvist15} consists of ultraslow neutrons that scatter 
on a very long current-carrying straight wire. The required low speed of the neutrons (typically 
in the order of cm/s) in combination with high angular resolution and precise centering 
of the beam is experimentally challenging. In order to deal with this, we develop here an 
interferometry-based experimentally simpler version of the scattering setup. We demonstrate 
how the Berry phase of the neutron spin associated with the magnetic field circling around 
the wire can be tested for feasible electrical currents and existing interferometry techniques 
for thermal neutrons in a triple-Laue setup \cite{rauch00}.   

\section{Adiabatic spin dynamics}

\begin{figure*}[ht]
\centering
\includegraphics[width=0.48\textwidth]{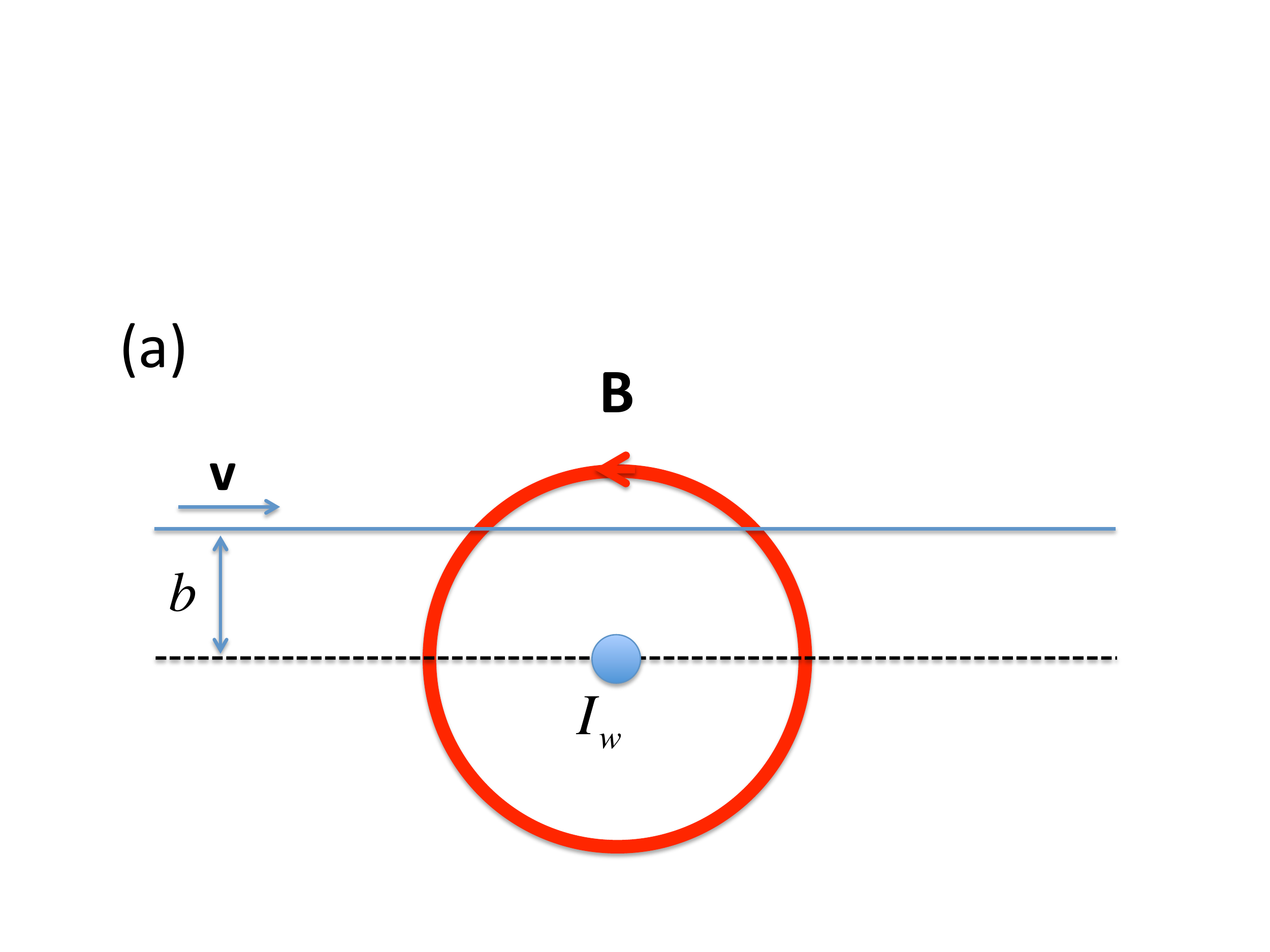}
\includegraphics[width=0.48\textwidth]{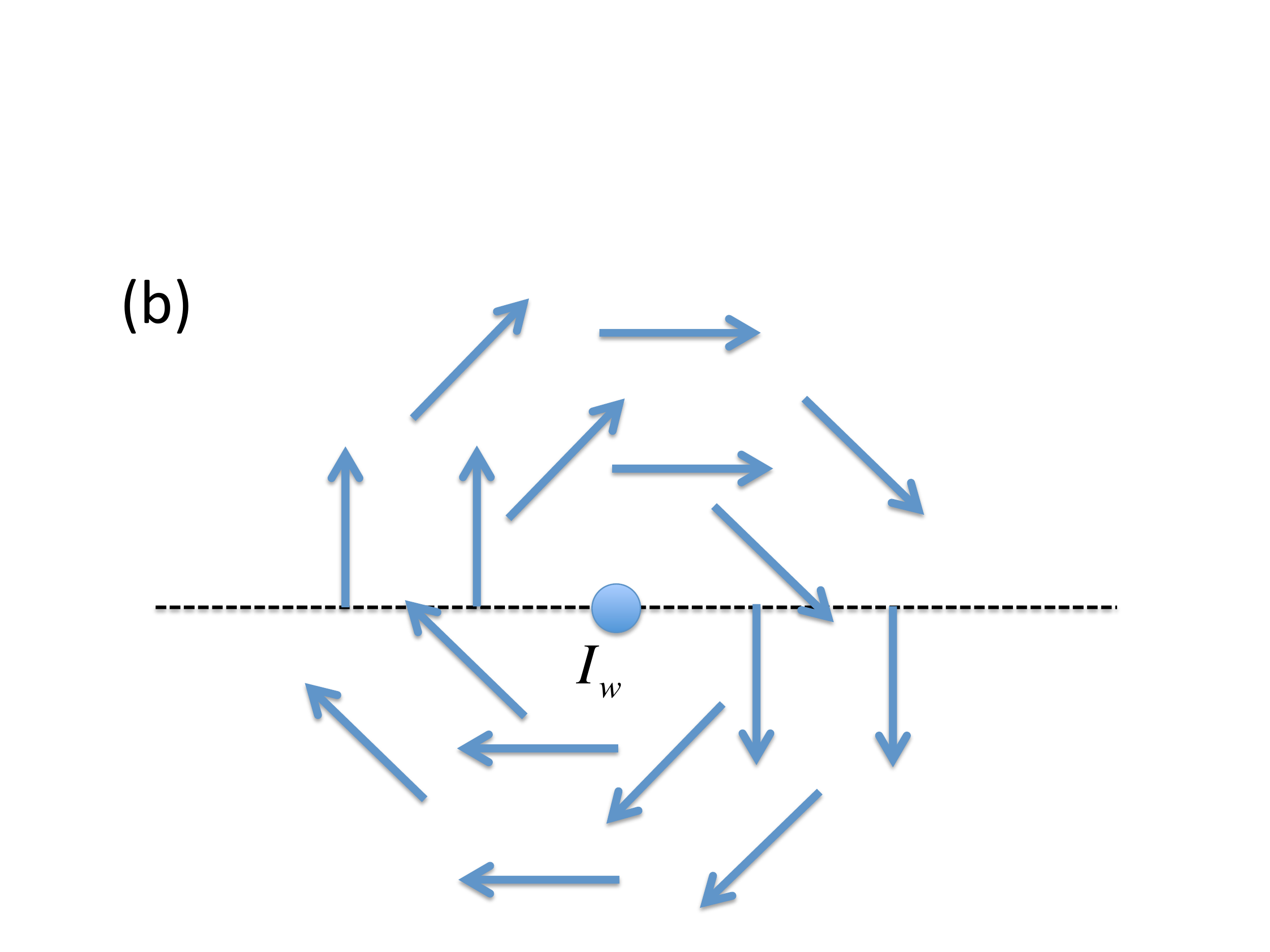}
\caption{(Color online) Left panel (a) shows a narrow beam of neutrons moving at velocity 
${\bf v}$ with an impact parameter $b$ on a very long straight wire carrying an electrical 
current $I_w$ perpendicular to the plane. Right panel (b) shows the local spin associated 
with the spin eigenstate $\ket{-;{\bf r}}$ (the local spin points in the opposite direction for 
$\ket{+;{\bf r}}$). The local spin is confined to the $xy$ plane with winding number $+1$. 
For sufficiently strong electrical current and sufficiently slow neutrons, the spin of the neutron 
follows adiabatically the circular magnetic field ${\bf B}$. This makes it possible to measure 
the Berry phase manifesting the nontrivial winding number associated with the circulating 
local spin.}
\label{fig:fig1}
\end{figure*}

Consider a narrow beam of neutrons moving at velocity ${\bf v} $ with an impact 
parameter $b$ on a very long straight wire carrying an electrical current $I_w$, as shown in 
Fig.~\ref{fig:fig1}a. The magnetic field is given by Biot-Savart's law 
\begin{eqnarray}
{\bf B} = \frac{\mu_0 I_w}{2\pi r} {\bf e}_{\theta} , 
\label{eq:bfield}
\end{eqnarray}
where we have assumed that the wire points along the $z$-axis. Here, $\theta$ is the 
polar angle in cylindrical coordinates with corresponding unit basis vector ${\bf e}_{\theta}$, 
$r$ is the distance from the wire, and $\mu_0 = 4\pi \cdot 10^{-7} \ {\rm{Vs}}/
{\rm{(Am)}}$ is permeability of vacuum. The magnetic field induces 
a local energy splitting of the neutron spin states, as described by the Zeeman 
Hamiltonian 
\begin{eqnarray} 
\mathcal{H} = - \boldsymbol{\mu} \cdot {\bf B} = - \mu \frac{1}{2} 
\boldsymbol{\sigma} \cdot {\bf B} 
\end{eqnarray} 
with $\mu = - 9.65 \cdot 10^{-27}$ J/T the 
neutron magnetic moment and  $\boldsymbol{\sigma}=\left( \sigma_x,\sigma_y,\sigma_z 
\right)$ the standard Pauli operators representing the neutron spin. The eigenvalues of the 
Hamiltonian are 
\begin{eqnarray} 
V_{\pm} (r) = \pm \frac{|\mu| \mu_0 I_w}{4\pi r} \equiv \pm \frac{C_0 I_w}{r} . 
\end{eqnarray} 
Numerically, we find $C_0 = 9.65 \cdot 10^{-34} {\rm{Jm}}  {\rm{A}}^{-1}$. The 
local spin eigenvectors $\ket{\pm;{\bf r}}$ are eigenvectors of $\sigma_{\theta} = 
{\bf e}_{\theta} \cdot \boldsymbol{\sigma}$. Each eigenvector defines a local spin 
$\bra{\pm ; {\bf r}} \boldsymbol{\sigma} \ket{\pm ; {\bf r}}$ confined to the $xy$ 
plane with winding number $+1$, as shown in Fig.~\ref{fig:fig1}b. A Berry phase 
factor of $-1$ when the neutron circles adiabatically around the wire once is a 
manifestation of the nontrivial topology of this circulating planar spin. 

In a semi-classical approach, the spatial motion of the neutrons is treated classically, 
while their spin is a quantum-mechanical degree of freedom. The orbits associated with 
the two spin eigenstates are solutions of 
\begin{eqnarray}
m_n\dot{{\bf v}} = -\nabla_{\bf r} V_{\pm} (r) = \pm \frac{C_0 I_w}{r^2} {\bf e}_r ,
\end{eqnarray}
where $m_n$ is the mass of the neutron. By using that the mechanical energy 
\begin{eqnarray}
E = \frac{m_n}{2} \dot{r}^2 + \frac{L^2}{2m_nr^2} \pm \frac{C_0 I_w}{r}  
\end{eqnarray}
and the angular momentum along the $z$ axis 
\begin{eqnarray}
L = m_n r^2 \dot{\theta} 
\end{eqnarray}
are constants of the motion, we find the solution 
\begin{eqnarray}
r = \frac{L^2}{m_n C_0 I_w} \left( 1 + \sqrt{1+\frac{2EL^2}{m_n C_0^2 I_w^2}} 
\sin \theta \right)^{-1} . 
\end{eqnarray}
Here, we have assumed that the neutron initially moves horizontally from the left with the 
origin at the center of the wire, as shown in Fig.~\ref{fig:fig1}a. In the interferometer 
setup, $E>0$ so that the eccentricity
\begin{eqnarray}
\varepsilon = \sqrt{1+\frac{2EL^2}{m_n C_0^2 I_w^2}} > 1 . 
\label{eq:eccentricity}
\end{eqnarray}
Thus, the trajectory is hyperbolic. 

Next, we analyze the adiabatic condition for the Zeeman Hamiltonian of the 
quantum-mechanical spin interacting with the magnetic field along the orbit 
of the neutron. In the adiabatic regime, the local spin is parallel or anti-parallel 
to the magnetic field at the location of the neutron. We examine the adiabatic 
condition for the spin evolution as governed by the semi-classical Hamiltonian 
\begin{eqnarray}
\mathcal{H}_{sc} = - \mu \frac{1}{2} \boldsymbol{\sigma} \cdot {\bf B} ({\bf r}_t) . 
\end{eqnarray}
In other words, for the local spin eigenstates $\ket{\pm;{\bf r}_t}$ we check whether 
or not the adiabatic condition 
\begin{eqnarray} 
\hbar \left| \frac{\bra{+;{\bf r}_t} \dot{\mathcal{H}}_{sc} ({\bf r}_t) 
\ket{-;{\bf r}_t}}{(E_{+} - E_{-})^2} \right| \ll 1
\label{eq:condition1}
\end{eqnarray}
is satisfied. We find: 
\begin{eqnarray}
\dot{\mathcal{H}}_{sc} ({\bf r}_t) = -\frac{C_0 I_w}{r^2} \left( v_r \sigma_{\theta} + 
v_{\theta} \sigma_r \right) , 
\end{eqnarray}
where $v_r = \dot{r}$ ($v_{\theta} = r\dot{\theta}$) is the radial (angular) velocity and 
$\sigma_r = {\bf e}_r \cdot \boldsymbol{\sigma}$, ${\bf e}_r$ being a unit basis vector 
in the radial direction. Since $\ket{\pm;{\bf r}_t}$ are eigenvectors of $\sigma_{\theta}$, 
it follows that $\bra{+;{\bf r}_t} \sigma_{\theta} \ket{-;{\bf r}_t} = 0$ and 
\begin{eqnarray}
\bra{+;{\bf r}_t} \dot{\mathcal{H}}_{sc} ({\bf r}_t) \ket{-;{\bf r}_t} & = &  
-\frac{C_0 I_w}{r^2} v_{\theta} \bra{+;{\bf r}_t} \sigma_r \ket{-;{\bf r}_t} 
\nonumber \\ 
 & = & i\frac{C_0 I_w}{r^2} v_{\theta}  . 
\label{eq:numerator}
\end{eqnarray}
Furthermore, 
\begin{eqnarray}
\left( E_{+} - E_{-} \right)^2 = \frac{4C_0^2 I_w^2}{r^2} .
\label{eq:denominator}
\end{eqnarray}
By inserting Eqs.~(\ref{eq:numerator}) and (\ref{eq:denominator}) into Eq.~(\ref{eq:condition1}), 
we finally obtain the adiabatic condition 
\begin{eqnarray}
\hbar \frac{C_0 I_w}{r^2} \left| v_{\theta} \right| \frac{r^2}{4C_0^2 I_w^2} & = & 
\frac{\hbar}{4 C_0 I_w} \left| v_{\theta} \right| \ll 1 
\nonumber \\ 
 & \Rightarrow & I_w \gg \frac{\hbar}{4 C_0} \left| v_{\theta} \right| . 
\label{eq:condition2}
\end{eqnarray} 

\section{Interferometer setup}

\begin{figure*}[ht]
\centering
\includegraphics[width=0.48\textwidth]{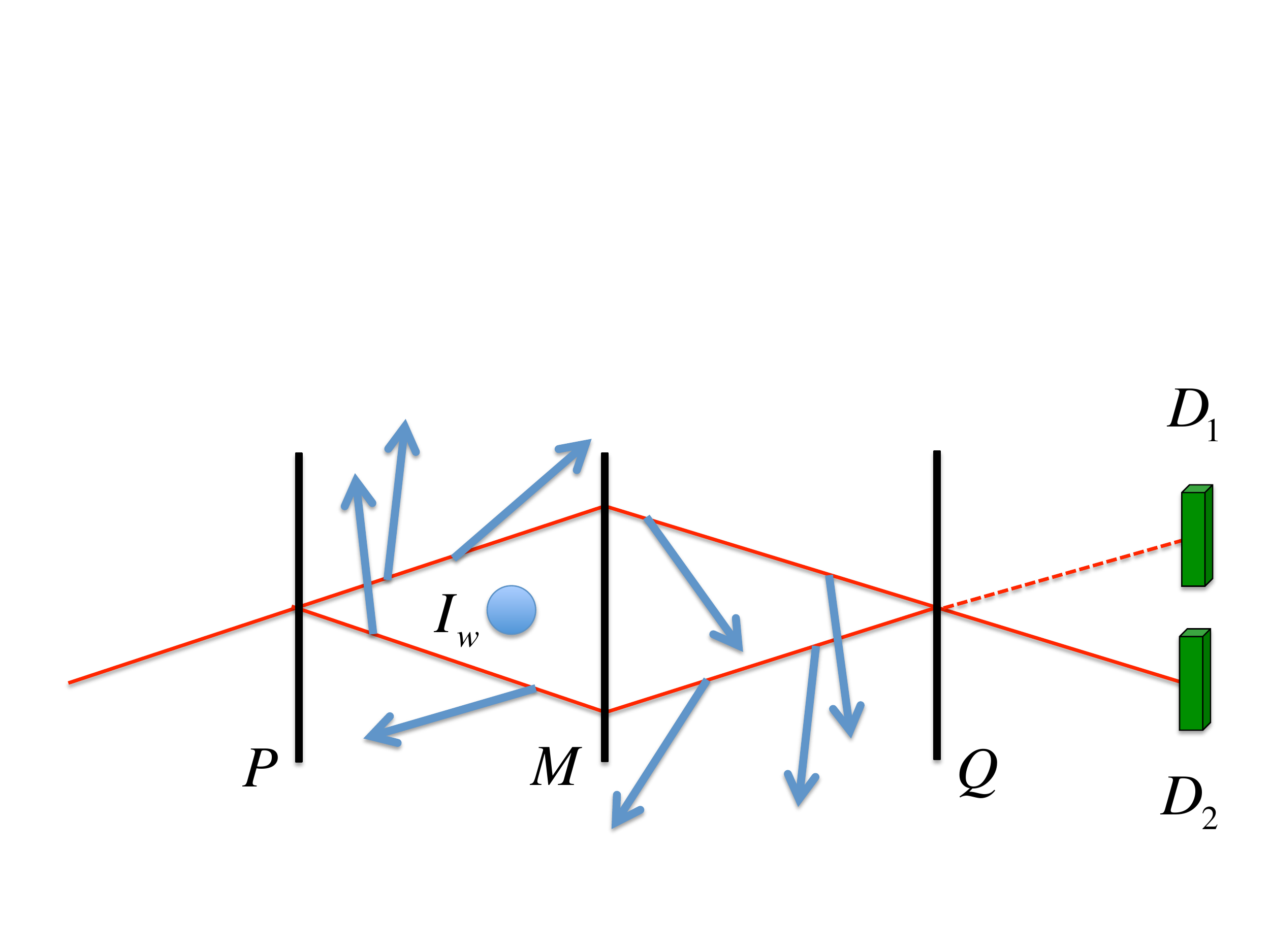}
\caption{(Color online) Triple-Laue interferometer setup to test the Berry phase associated 
with the local spin around the interferometer loop. A beam of thermal neutrons is coherently 
superposed at the first crystal plate $P$. The second crystal plate $M$ acting as mirrors 
recombine the beam-pair at the third crystal plate $Q$. The output intensity is measured at 
the two detectors $D_1$ and $D_2$. In order to cancel unwanted dynamical phase effects, the 
current-carrying wire is located on  the symmetry line connecting the two beam crossing points 
at $P$ and $Q$. The Berry phase associated with the planar spin gives rise to a destructive 
interference in the direction of the incoming beam so that all neutrons are detected at $D_2$.}
\label{fig:fig2}
\end{figure*}

Thermal neutrons moving at speed $v \sim 2000$ m/s have been used extensively for interferometry 
tests \cite{rauch00}. We demonstrate that such neutrons can be used to observed the above 
described Berry phase of the circulating planar spin. To this end, we first show that there exists 
a wide electrical current range for which the neutron moves along a straight line, and its spin 
evolution at the same time satisfies the adiabatic condition in Eq.~(\ref{eq:condition2}). 

We use that $E=\frac{1}{2}m_nv^2$ and $L=m_nvb$, $b$ being the impact parameter (see 
Fig.~\ref{fig:fig1}a), are constants of the motion, to rewrite the eccentricity as 
\begin{eqnarray}
\varepsilon = \sqrt{1+\frac{m_n^2 v^4 b^2}{C_0^2 I_w^2}} . 
\label{eq:eccentricity}
\end{eqnarray}
The neutron moves along a straight line provided $\varepsilon \gg 1$, which combined with 
Eq.~(\ref{eq:condition2}) yields 
\begin{eqnarray}
\frac{\hbar}{4 C_0} \left| v_{\theta} \right| \ll I_w \ll \frac{m_n v^2 b}{C_0} . 
\end{eqnarray}
For an impact parameter in the order of $10^{-1}$ m and $\max \left( \left| v_{\theta} \right| \right)
\sim v \sim 2000$ m/s, we find 
\begin{eqnarray}
50 \rm{A} \ll I_w \ll 7 \cdot 10^{11} \rm{A} . 
\end{eqnarray} 
This confirms that the straight line assumption can be combined with adiabatic spin evolution.  
By taking a reasonable upper limit for current density of 500 A/cm$^2$, a Copper wire of 
radius $0.5$ cm would allow for a sufficiently large current in the order of $400$ A.

The triple-Laue interferometer setup designed to measure the Berry phase is shown in 
Fig.~\ref{fig:fig2}. A beam of thermal neutrons is coherently superposed at the first 
crystal plate $P$ acting as a beam-splitter in front of the wire. A second crystal plate 
$M$ acting as mirrors recombine the beam-pair at the third crystal plate $Q$. The intensity 
at the two output detectors $D_1$ and $D_2$ is finally measured. At $Q$, the probability amplitude  
reads 
\begin{eqnarray}
\psi_Q & = & \frac{1}{\sqrt{2}} \psi_{\rm{up}} 
e^{i\int_{\rm{up};P}^Q  {\bf A}_{\pm} ({\bf r}) \cdot d{\bf r}} 
\nonumber \\ 
 & & + \frac{1}{\sqrt{2}} \psi_{\rm{down}} 
e^{i\int_{\rm{down};P}^Q  {\bf A}_{\pm} ({\bf r}) \cdot d{\bf r}} , 
\end{eqnarray}
where 
\begin{eqnarray}
{\bf A}_{\pm} ({\bf r}) = i\bra{\pm;{\bf r}} \nabla_{{\bf r}} \ket{\pm;{\bf r}} = 
-\frac{1}{2r} {\bf e}_{\theta}
\label{eq:berrypotential}
\end{eqnarray}
are the Berry vector potentials of the two spin eigenstates and $\psi_{\rm{up}}$ 
($\psi_{\rm{down}}$) is the probability amplitude for the upper (lower) beam. The vector 
potential is of Aharonov-Bohm type and corresponds to half a flux unit 
at the wire \cite{remark2}. The beam-splitter at $Q$ induces the transformations $\psi_{\rm{up}} 
\mapsto \psi_{D_1} + \psi_{D_2}$ and $\psi_{\rm{down}} \mapsto \psi_{D_1} - 
\psi_{D_2}$ yielding the output 
\begin{eqnarray}
\psi_{\rm{out}} & = & \frac{1}{2} 
\psi_{D_1} e^{i\int_{\rm{up};P}^Q  {\bf A}_{\pm} ({\bf r}) \cdot d{\bf r}} 
\left( 1 + e^{i\oint_C  {\bf A}_{\pm} ({\bf r}) \cdot d{\bf r}} \right) 
\nonumber \\ 
 & & + \frac{1}{2} \psi_{D_2} e^{i\int_{\rm{up};P}^Q  {\bf A}_{\pm} ({\bf r}) \cdot d{\bf r}} \left( 1 - 
e^{i\oint_C  {\bf A}_{\pm} ({\bf r}) \cdot d{\bf r}} \right) , 
\nonumber \\ 
\end{eqnarray}
where $C$ is the interferometer loop in the counterclockwise direction and $\psi_{D_1}$ 
($\psi_{D_2}$) is the amplitude for the upper (lower) output beam. The gauge independent 
$e^{i\oint_C  {\bf A}_{\pm} ({\bf r}) \cdot d{\bf r}}$ is the Berry phase factor associated with 
the planar local spin of the neutron. By using Eq.~(\ref{eq:berrypotential}), we 
find $e^{i\oint_C  {\bf A}_{\pm} ({\bf r}) \cdot d{\bf r}} = e^{i\pi} = -1$, yielding 
\begin{eqnarray}
\psi_{\rm{out}} & = & 
\psi_{D_2} e^{i\int_{\rm{up};P}^Q  {\bf A}_{\pm} ({\bf r}) \cdot d{\bf r}} . 
\end{eqnarray}
This shows that the Berry phase of the planar spin gives rise to a destructive interference 
effect that surpresses the probablity amplitude in the direction of the incoming 
beam, leading to that all neutrons are detected at $D_2$. This is the interferometer analog 
of the destructive interference effect that creates a nodal line in the forward direction seen in 
the scattering setup proposed in Ref.~\cite{sjoqvist15}. 

In general, a dynamical phase is accompanying the Berry phase. This phase may result in 
a dynamical contribution that would be sensitive to fluctuations in the neutron speed and 
thereby potentially destabilize the interference effect. This can be resolved by putting the 
wire on the symmetry line of the interferometer connecting the two beam crossing points at 
$P$ and $Q$. In this way, the dynamical phase contributions along the two beams are 
identical and therefore cancel so that only the Berry phase influences the interference effect 
measured at the two detectors $D_1$ and $D_2$. 

A typical neutron source produces unpolarized neutrons, as characterized by a spin 
density operator $\rho = \frac{1}{2} \hat{1}$. The interference effect is determined by 
the time evolution operator 
\begin{eqnarray}
U(Q;P) & = & e^{i\oint_C {\bf A}_{+} ({\bf r}) \cdot d{\bf r}} 
\ket{+} \bra{+} 
\nonumber \\ 
 & & + e^{i\oint_C {\bf A}_{-} ({\bf r}) \cdot d{\bf r}} \ket{-} \bra{-} = -\hat{1}
\end{eqnarray} 
of the spin, yielding the interference effect \cite{sjoqvist00} 
\begin{eqnarray}
I_{D_1} & = & I_{\rm{in}} - I_{D_2} 
\nonumber \\ 
 & \propto &  
1 + \left| \Tr (\rho U) \right| \cos \left[ \arg \Tr (\rho U) \right] = 0 . 
\end{eqnarray}
Thus, unpolarized neutrons can be used to demonstrate the Berry phase induced 
desctructive interference effect. This considerably simplifies the experimental realization 
of the setup as the neutrons can be used directly from the source without undergoing 
any intensity-reducing spin filtering. 

\section{Conclusions}
A very long straight wire carrying an electrical current defines a circulating planar local spin 
of a neutron. We have shown that there exists an electrical current range for which sufficiently 
slow neutrons can be used to test the Berry phase accompanying adiabatic evolution of 
the local spin around the interferometer loop. A complete cancelation of the 
probability amplitude in the direction of the incoming neutron beam provides a clear signature 
of the topological Berry phase of the circulating planar spin. The adiabatic condition is satisfied 
for electrical currents in the order of a few hundreds of A in the case of thermal neutrons 
in a triple-Laue setup. 

The Aharonov-Casher effect \cite{aharonov84} occurs when an electrical neutral particle 
moves around a charged straight wire. The resulting phase shift is topological in that it 
only depends on the winding number of the particles path around the wire. The effect 
has been observed by using thermal neutrons \cite{cimmino89}. Our poposed experiment 
can be viewed as an analog to this effect where the static charges are replaced be moving 
charges that constitute the electrical current through the uncharged wire. We note that, 
while the Aharonov-Casher phase shift is tiny (it is essentially a relativistic effect), the 
proposed Berry phase shift is large (integer multiple of $\pi$) and can therefore easily 
be observed provided the adiabatic condition is satisfied and the dynamical phases can 
be made to vanish by aligning the interferometer setup. 

Our proposed experiment is built upon existing well-proven techniques and requires 
achievable experimental parameter values. On the other hand, the corresponding 
scattering experiment, put forward in Ref.~\cite{sjoqvist15} and designed to test the Berry 
phase of the planar spin, requires ultraslow neutrons moving at a speed which is four 
to five orders of magnitude lower than for thermal neutrons. In combination with 
the high angular resolution and precise centering of the neutron beam, the need 
for such slow neutrons would be highly challenging. The present neutron interferometry 
setup avoids these problems and opens up for a simpler verification of the Berry phase 
of the circulating planar spin.  
\vskip 0.3 cm
Support from the Swedish Research Council Grant No. D0413201 is acknowledged.


\begin{thebibliography}{99}
\bibitem{remark2} In a more fine-grained description, the two spin states must cross energetically 
at a point inside the wire where the magnetic field vanishes. Indeed, if we assume a homogeneous 
electrical current density, Biot-Savart's law entails that the magnetic field tends to zero towards 
the center of the wire thereby defining the precise location of the Berry flux.
\end{thebibliography}
\end{document}